\documentclass[preprint,showpacs,nofootinbib]{revtex4}
\usepackage{graphicx}
\usepackage{epsfig,wick}

\newcommand{\beq}{\begin{equation}}
\newcommand{\eeq}{\end{equation}}
\newcommand{\bea}{\begin{eqnarray}} 
\newcommand{\eea}{\end{eqnarray}}

\renewcommand{\a}{\alpha}
\renewcommand{\b}{\beta}
\renewcommand{\c}{\gamma}
\renewcommand{\d}{\delta}
\newcommand{\e}{\epsilon}

\newcommand{\pa}{\partial}

\newcommand{\nn}{\nonumber\\ }

\begin{document}

\renewcommand{\thefootnote}{\fnsymbol{footnote}}

\title{PP-wave GS superstring, polygon divergent 
structure and conformal field theory}

\author{Mark A. Walton\footnote{walton@uleth.ca} and 
Jian-Ge Zhou\footnote{jiange.zhou@uleth.ca}}
\affiliation{ Physics Department, University of Lethbridge,
Lethbridge, Alberta, Canada T1K 3M4}

\begin{abstract}
In the semi-light cone
gauge  $g_{ab}=e^{2\phi}\delta_{ab}$,
$\bar{\gamma}^+\theta=0$,
we evaluate the $\phi$-dependent
effective action for the pp-wave Green-Schwarz (GS) superstring in 
both harmonic and group coordinates.
When we compute the fermionic $\phi$-dependent effective action
in harmonic coordinates, 
we find a new triangular one-loop Feynman diagram.
We show that the bosonic
$\phi$-dependent effective action cancels with the fermionic one,
indicating that the pp-wave GS superstring is a
conformal field theory. 
We introduce the group coordinates preserving $SO(4)\times\,SO(4)$
and conformal symmetry. Group coordinates are interesting because
vertex operators take simple forms in them.
The new feature in group coordinates is that there are
logarithmic divergences from n-gons, so that the divergent structure
is more complicated than in harmonic coordinates.
After summing over all contributions from n-gons, we show that
in group coordinates, the GS superstring on pp-wave RR 
background is still a conformal field theory.
\end{abstract}

\maketitle

\renewcommand{\thefootnote}{\arabic{footnote}}

\section{Introduction}
\label{s1}

Recently, Berenstein, Maldacena, and Nastase (BMN) \cite{BMN} argued that
the states of type IIB string theory on a plane-wave (pp-wave)
Ramond-Ramond (RR) background \cite{pprr} correspond to 
a class of operators in ${\cal N}=4$ $SU(N)$ Yang-Mills
with large $R$-charge $J$.
The type IIB string worldsheet action on the pp-wave RR background was
constructed using the Green-Schwarz (GS) formalism in \cite{mets, metst}.
In the light-cone gauge, the exact solvability of the Green-Schwarz 
string theory on pp-wave RR background provides for an explicit
realization of the pp-wave/SYM correspondence.

Although this light-cone formulation is very useful in calculating
the string spectrum, it is not manifestly superconformally
invariant.  This makes it difficult to compute scattering
amplitudes. Some covariant versions of type IIB string theory
on pp-wave RR background were proposed \cite{berk, berkm}
by exploiting the hybrid and $U(4)$ formalisms, but since
the most straightforward method for constructing
the superstring action on pp-wave RR background is the
Green-Schwarz formalism \cite{mets, metst}, 
it is interesting to check the conformal invariance of the pp-wave
GS superstring action explicitly.
To calculate pp-wave string interaction in 
the context of the perturbative
string theory, besides the conformal invariant action, one
has to construct string vertex operators. In a pp-wave 
background, the vertex operators
take a quite complicated form in harmonic coordinates,\footnote{The
harmonic coordinates are convenient in calculating
the string spectrum, but awkward for computing scattering
amplitudes. Group coordinates have the opposite properties.} but
they are simplified in group coordinates \cite{gv, jn}.
As the transformation between harmonic coordinates and group coordinates
is highly nonlinear, if we treat group coordinates as
``fundamental fields'' in the calculation, it is not
{\sl a priori} obvious whether
the pp-wave GS superstring action in  group coordinates is superconformal
invariant or not.

In general backgrounds, because of the mixing of vertex operators,
if one constructs the vertex operators like 
$G_{\mu\nu}\pa^{n}x^{\mu}{\bar{\pa}}^{n}x^{\nu}$, one has to
calculate n-loop Feynman diagrams for arbitrary n \cite{cg}.
This is  extremely
difficult to carry out in practice. If we compute the
pp-wave string interaction including BMN operators, it seems
some n-loop Feynman diagrams have to be involved. Thus it is
interesting to study the structure of the n-loop Feynman diagrams
of the GS superstring theory on pp-wave RR background.

In this paper, we study the GS superstring in the semi-light cone
gauge  $g_{ab}=e^{2\phi}\delta_{ab}$,
$\bar{\gamma}^+\theta=0$
on pp-wave RR background and its conformal invariance.
Following Kallosh-Morozov \cite{km},   
we calculate the partition function of the pp-wave
GS superstring in the semi-light cone gauge, 
and rewrite the original GS superstring
action with its $SO(8)$ spinors in a concise form using
$SU(4)$ spinors.  When $m=0$, the conformal anomaly from $SU(4)$ spinors
has a coefficient $+8$ while the $x^{\mu}$
contribute $\frac{10}{2}$ and reparametrization ghost contribute
$-\frac{26}{2}$, thus the total conformal anomaly 
is $\frac{10}{2} - \frac{26}{2} + 8 = 0$.

When $m\neq 0$, we first evaluate the $\phi$-dependent bosonic
effective action in harmonic coordinates.
Then we compute the fermionic $\phi$-dependent effective action, 
we find a new triangular one-loop Feynman diagram.
We show that the bosonic
$\phi$-dependent effective action cancels with the fermionic one, 
which indicates that the pp-wave GS superstring is an
exact conformal field theory. 
The quartic interaction term in pp-wave background
takes the special form ${x^{i}}^{2}\pa_{+}x^{+}\pa_{-}x^{+}$.
Since 
$x^{+}$ can only contract with $x^{-}$, there is no interacting
term quadratic in $x^{-}$, thus the higher-loop Feynman
diagrams can be decomposed as the products of the one-loop diagrams,
which is an important feature when we discuss the mixing
of the vertex operators.

We introduce the group coordinates preserving $SO(4)\times\,SO(4)$
and conformal symmetry. The new feature in group coordinates is that there are
logarithmic divergences from n-gons whose divergent structure
is more complicated than that in harmonic coordinates.
After summing over all contributions from n-gons, however, we find that
in group coordinates, the GS superstring on pp-wave RR 
background is still a conformal field theory.

The paper is organized as follows. In Section 2, we discuss the 
partition function of the GS superstring on the pp-wave RR 
background. In Section 3, we calculate the $\phi$-dependent
effective action and quantum counterterms in harmonic
coordinates, and find a new fermionic triangular one-loop
diagram. In Section 4, we study the quantum counterterm
and the $\phi$-dependent effective action in group coordinates.
Some interesting logarithmic divergences from n-gons
are found, which is a peculiar feature of the group coordinates.
In Section 5, we present our summary and conclusion.

\section{Partition function of GS superstring on pp-wave RR 
background}
\label{s2}

Let us start with the ten dimensional plane wave space with the
metric and RR-flux \cite{pprr}
\bea
ds^2\,&=& 2dx^+\,dx^-\,-m^2\,{x^i}^2\,d{x^+}^2 + d{x^i}^2\,,\nn
F_{+1234}&=&F_{+5678}=2m
\label{metric}
\eea
where $i=1,\cdot\cdot\cdot,8$, and $x^{\pm}=(x^9\pm\,x^0)/\sqrt{2}$.
The coordinates $x^{\pm}$, $x^i$ are called harmonic coordinates
and are physically convenient in calculating the string spectrum \cite{gv, jn}.

The $\kappa$-symmetry gauge fixed type IIB GS
superstring Lagrangian in the pp-wave RR background (\ref{metric}) with 
$\bar{\gamma}^+\theta=\bar{\gamma}^+{\bar\theta}=0$\footnote{We adopt
the same notation as in \cite{mets}, where the coset superspace \cite{zhou}
has been exploited.} is \cite{mets}
\bea
{\cal\, L}&=&-\frac{1}{2}\sqrt{g}g^{ab}(2\pa_{a}x^{+}\pa_{b}x^{-}-
m^{2}{x^{i}}^{2}\pa_{a}x^{+}\pa_{b}x^{+} + 
\pa_{a}x^{i}\pa_{b}x^{i})\nn
&&
-i\sqrt{g}g^{ab}\pa_{a}x^{+}(\bar{\theta}\bar{\gamma}^-\,\pa_{b}\theta
+ \theta\bar{\gamma}^-\,\pa_{b}\bar{\theta}\, + 2\,i\,m\pa_{b}x^{+}
{\bar\theta}{\bar\gamma}^-\,\Pi\theta)\nn
&&
+ i\e^{ab}\pa_{a}x^{+}(\theta\bar{\gamma}^-\,\pa_{b}\theta
+ \bar{\theta}\bar{\gamma}^-\,\pa_{b}\bar{\theta})
\eea
where $\Pi=\c^1\bar{\c}^2\c^3\bar{\c}^4$,  
$\theta=(\theta^1\,+i\theta^2)/\sqrt{2}$,
$\bar{\theta}=(\theta^1\,-i\theta^2)/\sqrt{2}$, $\theta^1$ and  $\theta^2$
are real Majorana-Weyl spinors or $SO(8)$ spinors. 

The partition function in the path integral formalism can
be computed by fixing the semi-light cone gauge, i.e., 
$\bar{\gamma}^+\theta=\bar{\gamma}^+{\bar\theta}=0$ and 
$g_{ab}=e^{2\phi}\delta_{ab}$. The gauge fixed path integral takes 
the form
\beq
Z=\int\,Dx^{\mu}D\theta^{1}D\theta^{2}DbDc\, (\det\pa_{+}x^{+})^{-4}
(\det\pa_{-}x^{+})^{-4}\,e^{-S}
\eeq
with
\bea
S&=&-\frac{1}{2\pi\a^{\prime}}\int\,d^{2}\sigma\,\bigg(
\pa_{+}x^{-}\pa_{-}x^{+}
+\frac{1}{2}\pa_{+}x^{i}\pa_{-}x^{i} - 
\frac{1}{2}m^{2}{x^{i}}^{2}\pa_{+}x^{+}\pa_{-}x^{+}\nn
&&
+ i\pa_{-}x^{+}\theta^1\,\bar{\gamma}^-\,\pa_{+}\theta^1\,
+ i\pa_{+}x^{+}\theta^2\,\bar{\gamma}^-\,\pa_{-}\theta^2\, -
2\,i\,m\pa_{+}x^{+}\pa_{-}x^{+}\theta^1\,{\bar\gamma}^-\,\Pi\theta^2
+ {\cal\, L}_{bc}\bigg)
\eea
where ${\cal\, L}_{bc}$ is the usual conformal ghost 
Lagrangian, and in Euclidean worldsheet we have $\pa_{\pm}=i\pa_{0}\pm\pa_{1}$.
The origin of  $(\det\pa_{+}x^{+})^{-4}(\det\pa_{-}x^{+})^{-4}$ can
be traced back to the Faddeev-Popov fermionic
gauge symmetry ghost Lagrangian 
$\sqrt{g}g^{+-}c^{(1)}\Gamma^{-}\Gamma^{+}\Gamma^{\mu}L^{+}_{+}\Gamma^{+}
\Gamma^{-}c^{(1)}_{-}$
$+$\\
$\sqrt{g}g^{-+}c^{(2)}\Gamma^{-}\Gamma^{+}\Gamma^{\mu}L^{+}_{-}\Gamma^{+}
\Gamma^{-}c^{(2)}_{+}$ $\rightarrow$ 
$\,(\det\pa_{+}x^{+})^{-8}(\det\pa_{-}x^{+})^{-8}$ and to 
$(\det\pa_{+}x^{+})^{4}(\det\pa_{-}x^{+})^{4}$ due to the existence
of second class constraints \cite{km}.

Using the fact that the $SO(8)$ spinor $\theta$ with
$\bar{\gamma}^+\theta=0$ can be presented as two  $SU(4)$ spinors
$\psi_{\a},\eta_{\a}, \a=1,\cdot\cdot\cdot,4$, the partition
function can be written as
\beq
Z=\int\,Dx^{\mu}D\psi^{1}D\eta^{1}D\psi^{2}D\eta^{2}DbDc\, 
(\det\pa_{+}x^{+})^{-4}
(\det\pa_{-}x^{+})^{-4}\,e^{-S}
\eeq
with
\bea
S&=&-\frac{1}{2\pi\a^{\prime}}\int\,d^{2}\sigma\,\bigg(
\pa_{+}x^{-}\pa_{-}x^{+}
+\frac{1}{2}\pa_{+}x^{i}\pa_{-}x^{i} - 
\frac{1}{2}m^{2}{x^{i}}^{2}\pa_{+}x^{+}\pa_{-}x^{+}\nn
&&
- i\pa_{-}x^{+}\psi^1\,\pa_{+}\eta^1\,
- i\pa_{+}x^{+}\psi^2\,\pa_{-}\eta^2\, 
+ i\,m\pa_{+}x^{+}\pa_{-}x^{+}(\psi^1\,\psi^2\, + \eta^1\eta^2\,)
+ {\cal\, L}_{bc}\bigg)
\eea
where we have exploited $\pa_{-}x^{+}\eta^1\pa_{+}\psi^1$ 
$\rightarrow$  $\pa_{-}x^{+}\psi^1\pa_{+}\eta^1$ 
$+$ $\psi^1\eta^1\pa_{+}\pa_{-}x^{+}$, 
$\pa_{+}x^{+}\eta^2\pa_{-}\psi^2$ 
$\rightarrow$  $\pa_{+}x^{+}\psi^2\pa_{-}\eta^2$ 
$+$ $\psi^2\eta^2\pa_{+}\pa_{-}x^{+}$, and
$x^{-}+\frac{i}{2}\psi^1\eta^1+\frac{i}{2}\psi^2\eta^2$
$\rightarrow$ $x^{-}$, that is, 
$(\frac{i}{2}\psi^1\eta^1+\frac{i}{2}\psi^2\eta^2)\pa_{+}\pa_{-}x^{+}$
can be absorbed into the redefinition of $x^{-}$.

After rescaling $\pa_{-}x^{+}\psi^1\rightarrow\,-\psi^1$
and $\pa_{+}x^{+}\psi^2\rightarrow\,-\psi^2$, we have
\beq
Z=\int\,Dx^{\mu}D\psi^{1}D\eta^{1}D\psi^{2}D\eta^{2}DbDc\,e^{-S}
\label{pf}
\eeq
with
\bea
S&=&-\frac{1}{2\pi\a^{\prime}}\int\,d^{2}\sigma\,\bigg(
\pa_{+}x^{-}\pa_{-}x^{+}
+ \frac{1}{2}\pa_{+}x^{i}\pa_{-}x^{i} 
+ i\psi^1\,\pa_{+}\eta^1\,+i\psi^2\,\pa_{-}\eta^2\, + {\cal\, L}_{bc}\nn
&&
-\frac{1}{2}m^{2}{x^{i}}^{2}\pa_{+}x^{+}\pa_{-}x^{+}
+ i\,m\psi^1\,\psi^2\,
+ i\,m\eta^1\eta^2\,\pa_{+}x^{+}\pa_{-}x^{+}\bigg)
\label{fs}
\eea
which consists of 10 ordinary scalar fields $x^{\mu}$,
reparametrization ghost $b,c$, and four $SU(4)$ fermions
$\psi^1_{\a},\eta^1_{\a},\psi^2_{\a},\eta^2_{\a}$
with $\a=1,\cdot\cdot\cdot,4$. 
The first line shows the kinetic terms, and the second line
is the interaction Lagrangian.
The action (\ref{fs}) is nothing but the
type IIB GS superstring action in the semi-light cone gauge on
the pp-wave RR background, where its original $SO(8)$ spinors
have been presented as $SU(4)$ spinors.
Here we point out that the action (\ref{fs}) is much simpler than
the superstring action derived using the $U(4)$
formalism in \cite{berkm}.

\section{Conformal invariance of GS superstring in harmonic coordinates}
\label{s3}

To show the action (\ref{fs}) is conformally invariant, let us
first consider the $m=0$ case. In (\ref{fs}), there are no worldsheet
fermions, but instead some $0$-differential
spacetime fermions $\eta^1,\eta^2$ 
and $1$-differential spacetime fermions $\psi^1,\psi^2$. The 
$1$-differential fermions $\psi^1$ do not have an ordinary
norm $\|\psi^1\|^{2}\,=\int\,d^2\sigma\sqrt{g}\psi^1\bar{\psi^1}$,
instead their norm is induced by that of $\theta^1$ and $\theta^2$.
From the rescaling $\pa_{-}x^{+}\psi^1\rightarrow\,-\psi^1$, the
appropriate norm for $\psi^1$ is \cite{km}
\beq
\|\psi^1\|^{2}\,=\int\,d^2\sigma\sqrt{g}\psi^1\bar{\psi^1}\frac{1}{
|\pa_{-}x^+|^2}
\eeq
and the regulator integral should be $\int|D\eta^1|^2|D\psi^1|^2\,
\exp\bigg(-\int\,d^2\sigma(\psi^1\pa_{+}\eta^1+\bar{\psi}^1\pa_{-}\bar{
\eta}^1+M^2\frac{\sqrt{g}}{|\pa_{-}x^+|^2}\psi^1\bar{\psi^1})\bigg)$.
Integrating out $\psi^1$, $\bar{\psi^1}$ fields gives 
$\int|D\eta^1|^2\,\exp\int\sqrt{g}\bar{\eta^1}\Delta\eta^1/M^2$
with Laplace operator of the form
\beq
\Delta = \frac{1}{\sqrt{g}}\pa_{-}\frac{|\pa_{-}x^+|^2}{\sqrt{g}}\pa_{+}.
\eeq
The conformal anomaly for the Laplace operator $\Delta_{p,q}=p(z)\pa\,
q(z)\bar{\pa}$ is described by Liouville action \cite{km}
\beq
\exp\bigg(-\frac{1}{48\pi}\int(\frac{|\pa\,p|^2}{p^2}-\frac{4\pa\,p
\bar{\pa\,q}}{pq}+\frac{|\pa\,q|^2}{q^2})\bigg)
\label{la}
\eeq
In conformal gauge $g_{ab}=e^{2\phi}\delta_{ab}$, i.e., $p=e^{-2\phi}$,
$q=e^{-2\phi}|\pa_{-}x^+|^2$, then (\ref{la}) becomes
\beq
\exp\bigg(-\frac{1}{48\pi}\int[-8\pa\phi\bar{\pa}\phi-
(4\phi+\ln|\pa_{-}x^+|^2)\pa\bar{\pa}\ln|\pa_{-}x^+|^2]\bigg).
\eeq
The first term in the bracket shows that the conformal anomaly from 
$\psi^1,\eta^1$ has a coefficient $+8$ while the $x^{\mu}$
contribute $\frac{10}{2}$ and reparametrization ghosts contribute
$-\frac{26}{2}$, so the total Polyakov anomaly in (\ref{fs}) with
$m = 0$ is $\frac{10}{2} - \frac{26}{2} + 8 = 0$. The second term in 
the bracket represents the additional anomaly, which is proportional
to $\pa\bar{\pa}x^+$ and can be removed by a shift of the $x^-$
field \cite{km}. For $\psi^2,\eta^2$, we have the same result.

When $m\neq\,0$, the dimensional continuation of the action (\ref{fs})
is
\bea
S&=&-\frac{1}{2\pi\a^{\prime}}\int\,d^{2+\epsilon}\sigma\,e^{\epsilon
\phi}\bigg(
\pa_{+}x^{-}\pa_{-}x^{+}
+ \frac{1}{2}\pa_{+}x^{i}\pa_{-}x^{i} 
+ i\psi^1\,\pa_{+}\eta^1\,+i\psi^2\,\pa_{-}\eta^2\, + {\cal\, L}_{bc}\nn
&&
-\frac{1}{2}m^{2}{x^{i}}^{2}\pa_{+}x^{+}\pa_{-}x^{+}
+ i\,m\psi^1\,\psi^2\,
+ i\,m\eta^1\eta^2\,\pa_{+}x^{+}\pa_{-}x^{+}\bigg)
\label{ds}
\eea
where we perform all the index algebra in two dimensions and continue
the volume element to $2+\epsilon$ dimensions \cite{ft,cfmp}. The trace of the
stress-energy tensor can be obtained by variation of the
effective action  with respect to $\phi$ \cite{cfmp}. 

If $m$ is small, we can calculate the $\phi$-dependence of the 
effective action by perturbation theory. Since in the nonlinear
sigma model the quadratic
divergences like the contact term $\d\,(0)$ can be omitted 
in dimensional regularization \cite{ft}, in the following calculation we
only consider
the logarithmical divergences.\footnote{Because of the equal
total numbers of bosons and fermions, the trivial quadratic
divergences cancel.}  When $\epsilon$ is small,
the action (\ref{ds}) can be written as
\bea
S&=&-\frac{1}{2\pi\a^{\prime}}\int\,d^{2+\epsilon}\sigma\,\bigg(
\pa_{+}x^{-}\pa_{-}x^{+}
+ \frac{1}{2}\pa_{+}x^{i}\pa_{-}x^{i} 
+ i\psi^1\,\pa_{+}\eta^1\,+i\psi^2\,\pa_{-}\eta^2\, + {\cal\, L}_{bc}\nn
&&
+\pa_{+}x^{-}\pa_{-}x^{+}\epsilon\phi +  
\frac{1}{2}\pa_{+}x^{i}\pa_{-}x^{i}\epsilon\phi
+i\psi^1\,\pa_{+}\eta^1\,\epsilon\phi
+i\psi^2\,\pa_{-}\eta^2\,\epsilon\phi
+ {\cal\, L}_{bc}\,\epsilon\phi\nn
&&
-\frac{1}{2}m^{2}{x^{i}}^{2}\pa_{+}x^{+}\pa_{-}x^{+}
-\frac{1}{2}m^{2}{x^{i}}^{2}\pa_{+}x^{+}\pa_{-}x^{+}\epsilon\phi
+ i\,m\psi^1\,\psi^2\,+ i\,m\psi^1\,\psi^2\,\epsilon\phi\nn
&&
+ i\,m\eta^1\eta^2\,\pa_{+}x^{+}\pa_{-}x^{+}
+ i\,m\eta^1\eta^2\,\pa_{+}x^{+}\pa_{-}x^{+}\epsilon\phi
\bigg)
\label{ps}
\eea
where the quadratic terms are usual kinetic terms, the
cubic and quartic terms are interacting ones. From (\ref{ps}), we
can read off the propagators as
\bea
\wick{1}{<1x^-\,(\sigma_{1})>1x^+\,(\sigma_{2})}&=&2\pi\a^{\prime}
\int\frac{d^{2+\epsilon}k}{(2\pi)^2}\frac{1}{k^2}e^{ik\cdot\,
(\sigma_{1}-\sigma_{2})}\nn
\wick{1}{<1x^i\,(\sigma_{1})>1x^j\,(\sigma_{2})}&=&2\pi\a^{\prime}
\delta^{ij}\int\frac{d^{2+\epsilon}k}{(2\pi)^2}\frac{1}{k^2}e^{ik\cdot\,
(\sigma_{1}-\sigma_{2})}\nn
\wick{1}{<1\psi_{\a}^1\,(\sigma_{1})>1\eta_{\b}^1\,(\sigma_{2})}&=&
2\pi\a^{\prime}
\delta_{\a\b}\int\frac{d^{2+\epsilon}k}{(2\pi)^2}\frac{1}{ik_{+}}e^{ik\cdot\,
(\sigma_{1}-\sigma_{2})}\nn
\wick{1}{<1\psi_{\a}^2\,(\sigma_{1})>1\eta_{\b}^2\,(\sigma_{2})}&=&
2\pi\a^{\prime}
\delta_{\a\b}\int\frac{d^{2+\epsilon}k}{(2\pi)^2}\frac{1}{ik_{-}}e^{ik\cdot\,
(\sigma_{1}-\sigma_{2})}
\label{pg}
\eea
where $k_{\pm}\,=k_{0}\mp\,ik_{1}$. For $x^{+}$ and $x^{-}$,
only the propagator $<x^{+}\,x^{-}>$ is nonzero, while averages of
any number of $x^{+}$ without $x^{-}$ vanish.

We first consider the bosonic $\phi$-dependent effective action,
that is, we switch off the fermionic fields. From (\ref{ps}), we find
that the first logarithmically divergent term
$B_{1}$  is given by 
\bea
B_{1}&=&\frac{1}{2\pi\a^{\prime}}\int\,d^{2+\epsilon}\sigma\,
\bigg(-\frac{1}{2}m^2\,\wick{1}{<1x^i\,>1x^i\,}
\pa_{+}x^{+}\pa_{-}x^{+}\epsilon\phi   
\bigg)\nn
&=&
-\frac{2m^2}{\pi}\int\,d^{2}\sigma\,\phi\,\pa_{+}x^{+}\pa_{-}x^{+}
\label{b1}
\eea
which is obtained by the contraction $<x^i\,x^i>$ and described by
Fig.1. In (\ref{b1}), the limit $\epsilon\rightarrow 0$ has been taken.

%%%%%%% Fig. 1 %%%%%%%%%%%%%%%%
\begin{figure}[htbp]
\begin{center}
\includegraphics*[scale=.60]{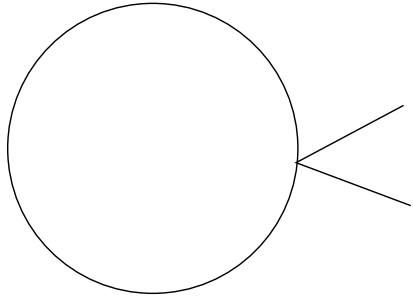}
\vspace{-0.7cm}
\end{center}
\caption{The one-loop logarithmically divergent term coming from $<x^i\,x^i>$.}
\end{figure}
%%%%%%%%%%%%%%%%%%%%%%%%%%%%%%%

The second one comes from the contraction between
the $\phi$-dependent kinetic term and the quartic term
\bea
B_{2}&=&\frac{2}{(2\pi\a^{\prime})^2}\int\,d^{2+\epsilon}\sigma_{1}\,
d^{2+\epsilon}\sigma_{2}\,
\frac{1}{2}\wick{12}{\pa_{+}<1x^i(\sigma_{1})\pa_{-}\,<2x^i(\sigma_{1})
\epsilon\phi(\sigma_{1})[-m^2\,>1x^j(\sigma_{2})>2x^j(\sigma_{2})}\nn
&&\cdot
\frac{1}{2}\pa_{+}x^+(\sigma_{2})\pa_{-}\,
x^+(\sigma_{2})]\nn
&=&
\frac{2m^2}{\pi}\int\,d^{2}\sigma\,\phi\,\pa_{+}x^{+}\pa_{-}x^{+}.
\label{b2}
\eea
The factor $2$ in the first line comes from two sorts of the
contraction between $\pa_{+}x^i\pa_{-}x^i$ and $x^j\,x^j$. The
Feynman diagram for $B_{2}$ is shown in Fig.2.

%%%%%%% Fig. 2 %%%%%%%%%%%%%%%%
\begin{figure}[htbp]
\begin{center}
\includegraphics*[scale=.60]{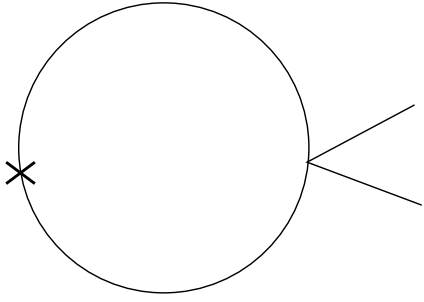}
\vspace{-0.7cm}
\end{center}
\caption{The logarithmic divergence coming from 
the contraction between
the $\phi$-dependent kinetic term and the quartic term, the
cross represents the insertion of $\epsilon\phi$.}
\end{figure}
%%%%%%%%%%%%%%%%%%%%%%%%%%%%%%%

The third logarithmically divergent term
$B_{3}$  is 
\bea
B_{3}&=&\frac{2}{(2\pi\a^{\prime})^2}\int\,
d^{2+\epsilon}\sigma_{1}\,
d^{2+\epsilon}\sigma_{2}\,
(-\frac{1}{2}m^2\,\wick{1}{<1x^i\,>1x^i\,}(\sigma_{1})
\pa_{+}x^{+}(\sigma_{1})\pa_{-}\wick{1}{<1x^{+}(\sigma_{1}) 
)\pa_{+}>1x^{-}}(\sigma_{2})\pa_{-}x^{+}(\sigma_{2})\epsilon\phi\nn
&=&
\frac{4m^2}{\pi}\int\,d^{2}\sigma\,\phi\,\pa_{+}x^{+}\pa_{-}x^{+}
\label{b3}
\eea
which is obtained by the contraction between the part of $B_{1}$
and $\pa_{+}x^{-}\pa_{-}x^{+}\epsilon\phi$, and can be described 
similarly by Fig.1.
We can check 
from the partition function and 
the action (\ref{ps}) that there are some higher-loop 
divergent contributions to the total bosonic
$\phi$-dependent effective action given by
\bea
B_{\phi}=\ln\,W-\frac{1}{2\pi\a^{\prime}}\int\,d^{2+\epsilon}\sigma\,
(-\frac{1}{2}m^{2}{x^{i}}^{2}\pa_{+}x^{+}\pa_{-}x^{+})
\label{b}
\eea
with
\bea
W &=&\lim_{\epsilon\rightarrow\,0}T\bigg\{\exp\bigg[\frac{1}{2\pi\a^{\prime}}
\int\,d^{2+\epsilon}\sigma\,(\pa_{+}x^{-}\pa_{-}x^{+}\epsilon\phi 
 +\frac{1}{2}\pa_{+}x^{i}\pa_{-}x^{i}\epsilon\phi
-\frac{1}{2}m^{2}{x^{i}}^{2}\pa_{+}x^{+}\pa_{-}x^{+}\nn
&&
-\frac{1}{2}m^{2}{x^{i}}^{2}\pa_{+}x^{+}\pa_{-}x^{+}\epsilon\phi
)\bigg]\bigg\}\nn
&=&\lim_{\epsilon\rightarrow\,0}\sum_{n=0}^{\infty}\frac{1}{n!}
\bigg[\frac{1}{2\pi\a^{\prime}}\int\,d^{2+\epsilon}\sigma\,
(-\frac{1}{2}m^2\,\wick{1}{<1x^i\,>1x^i\,}
\pa_{+}x^{+}\pa_{-}x^{+}\epsilon\phi) \bigg]^n\,
\sum_{k,l,s=0}^{\infty}\frac{1}{k!}\frac{1}{l!}\frac{1}{s!}
{k!}{s!}\left(
\begin{array}{c}
l\\
k
\end{array}
\right)
\left(
\begin{array}{c}
l-k\\
s
\end{array}
\right)\nn
&&
\bigg[\frac{2}{(2\pi\a^{\prime})^2}\int\,d^{2+\epsilon}\sigma_{1}\,
d^{2+\epsilon}\sigma_{2}\,
\frac{1}{2}\wick{12}{\pa_{+}<1x^i(\sigma_{1})\pa_{-}\,<2x^i(\sigma_{1})
\epsilon\phi(\sigma_{1})[-m^2\,>1x^j(\sigma_{2})>2x^j(\sigma_{2})}
\frac{1}{2}\pa_{+}x^+(\sigma_{2})\pa_{-}\,
x^+(\sigma_{2})]\bigg]^k\nn
&&\cdot
\bigg[\frac{2}{(2\pi\a^{\prime})^2}\int\,
,d^{2+\epsilon}\sigma_{1}\,
d^{2+\epsilon}\sigma_{2}\,
(-\frac{1}{2}m^2\,\wick{1}{<1x^i\,>1x^i\,}(\sigma_{1})
\pa_{+}x^{+}(\sigma_{1})\pa_{-}\wick{1}{<1x^{+}(\sigma_{1}) 
)\pa_{+}>1x^{-}}(\sigma_{2})\pa_{-}x^{+}(\sigma_{2})\epsilon\phi\bigg]^s\nn
&&\cdot
\bigg[\frac{1}{2\pi\a^{\prime}}
\int\,d^{2+\epsilon}\sigma\,(-\frac{1}{2}m^{2}{x^{i}}^{2}
\pa_{+}x^{+}\pa_{-}x^{+} )\bigg]^{l-k-s}
\label{w}
\eea
and the diagramatic illustration for the total bosonic
$\phi$-dependent effective action is drawn in Fig.3.
Fig.3 shows that the higher-loop diagrams can be decomposed
into the products of one-loop diagrams, which will be a crucial feature
when we discuss the mixing of the vertex operators on pp-wave RR background.

%%%%%%% Fig. 3 %%%%%%%%%%%%%%%%
\begin{figure}[htbp]
\begin{center}
\includegraphics*[scale=.60]{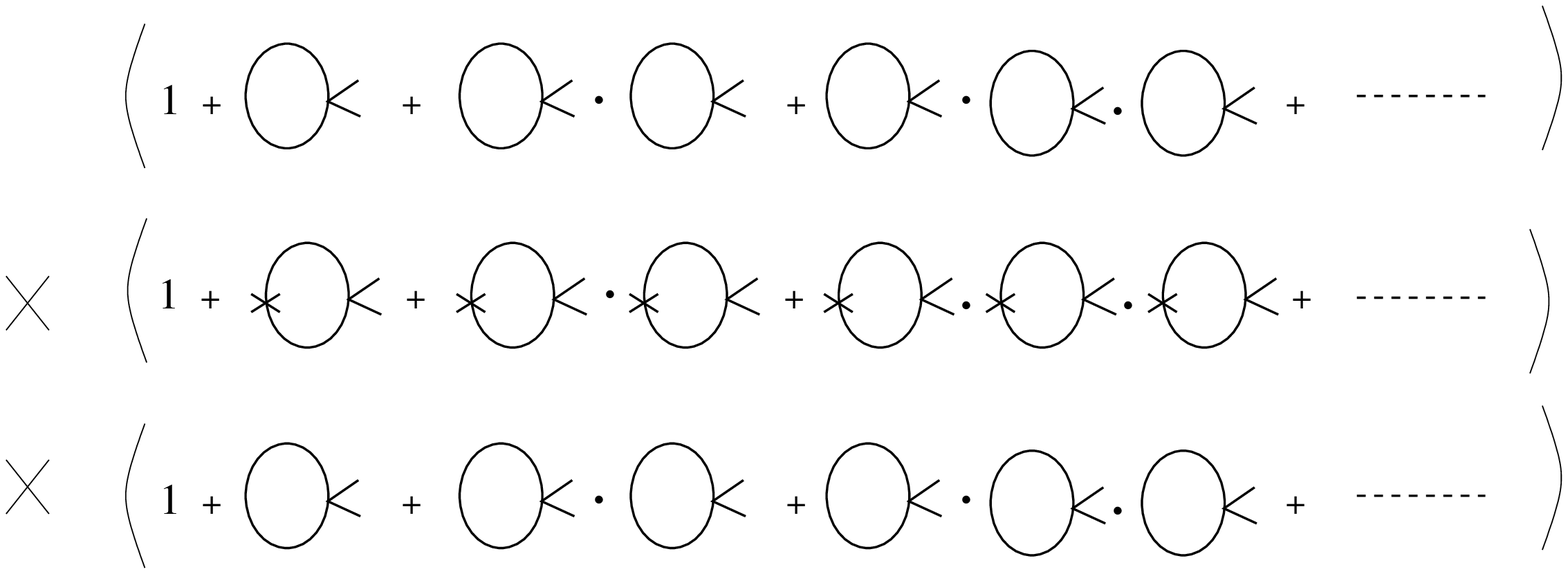}
\vspace{-0.7cm}
\end{center}
\caption{The total bosonic divergences from all loop contribution.}
\end{figure}
%%%%%%%%%%%%%%%%%%%%%%%%%%%%%%%

Inserting (\ref{b1}), (\ref{b2}) and  (\ref{b3}) into (\ref{w}), $W$
is reduced to
\bea
W &=&e^{-\frac{2m^2}{\pi}\int\,d^{2}\sigma\,\phi\,\pa_{+}x^{+}\pa_{-}x^{+}}
\cdot
e^{\frac{2m^2}{\pi}\int\,d^{2}\sigma\,\phi\,\pa_{+}x^{+}\pa_{-}x^{+}}\cdot
e^{\frac{4m^2}{\pi}\int\,d^{2}\sigma\,\phi\,\pa_{+}x^{+}\pa_{-}x^{+}}\nn
&&
\cdot
e^{\frac{1}{2\pi\a^{\prime}}
\int\,d^{2+\epsilon}\sigma\,(-\frac{1}{2}m^{2}{x^{i}}^{2}
\pa_{+}x^{+}\pa_{-}x^{+})}\nn
&=&
e^{\frac{4m^2}{\pi}\int\,d^{2}\sigma\,\phi\,\pa_{+}x^{+}\pa_{-}x^{+}}\cdot
e^{\frac{1}{2\pi\a^{\prime}}
\int\,d^{2+\epsilon}\sigma\,(-\frac{1}{2}m^{2}{x^{i}}^{2}
\pa_{+}x^{+}\pa_{-}x^{+})}.
\label{w1}
\eea
Plugging (\ref{w1}) into (\ref{b}), we have 
\bea
B_{\phi}=\frac{4m^2}{\pi}\int\,d^{2}\sigma\,\phi\,\pa_{+}x^{+}\pa_{-}x^{+}
\eea
which indicates that the total bosonic $\phi$-dependent effective action 
does not vanish on pp-wave background. In other words, the $\b$-function
 $\b_G$ is not zero \cite{gomez}.
It has been shown from other approaches that
the $D=26$ bosonic string on pp-wave 
defines a good string background \cite{lot} only when $\sum\limits_i\, A_i\,
=0$, but in our case it is $8m^2$.

We consider the fermionic contribution to the 
$\phi$-dependent effective action. The action (\ref{ps})
shows that there are two logarithmically divergent terms, the
first is
\bea
F_{1}&=&\frac{2}{(2\pi\a^{\prime})^2}\int\,d^{2+\epsilon}\sigma_{1}\,
d^{2+\epsilon}\sigma_{2}\,
im\wick{12}{<1\psi^1(\sigma_{1})<2\psi^2(\sigma_{1})
\epsilon\phi(\sigma_{1})im\,>1\eta^1\,(\sigma_{2})>2\eta^2(\sigma_{2})}\nn
&&\cdot
\pa_{+}x^+(\sigma_{2})\pa_{-}\,
x^+(\sigma_{2})\nn
&=&
\frac{4m^2}{\pi}\int\,d^{2}\sigma\,\phi\,\pa_{+}x^{+}\pa_{-}x^{+}
\label{f1}
\eea
where the factor $2$  in the first line comes from the contraction
between $im\psi^1\psi^2$ and $im\eta^1\,\eta^2\epsilon\phi$.
The Feynman diagram for $F_{1}$ is shown by Fig.4.

%%%%%%% Fig. 4 %%%%%%%%%%%%%%%%
\begin{figure}[htbp]
\begin{center}
\includegraphics*[scale=.60]{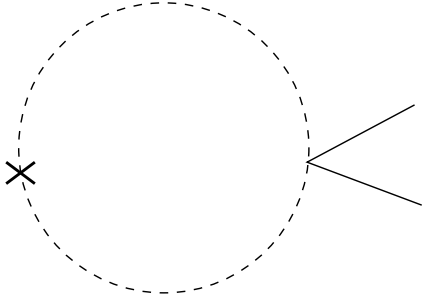}
\vspace{-0.7cm}
\end{center}
\caption{The fermionic one-loop logarithmic divergence from
the contraction between $\psi^1\psi^2$ and $\eta^1\eta^2$, where 
cross represents $\epsilon\phi$.}
\end{figure}
%%%%%%%%%%%%%%%%%%%%%%%%%%%%%%%

The second logarithmically divergent term is given by
\bea
F_{2}&=&\frac{2}{(2\pi\a^{\prime})^3}\int\,d^{2+\epsilon}\sigma_{1}\,
d^{2+\epsilon}\sigma_{2}\,d^{2+\epsilon}\sigma_{3}\,
i\wick{123}{<1\psi^1(\sigma_{1})\pa_{+}<2\eta^1(\sigma_{1})
\epsilon\phi(\sigma_{1})im\,>2\psi^1\,(\sigma_{2})<3\psi^2(\sigma_{2})
im>1\eta^1(\sigma_{3})>3\eta^2(\sigma_{3})}\nn
&&\cdot
\pa_{+}x^+(\sigma_{1})\pa_{-}\,
x^+(\sigma_{2})\nn
&=&
-\frac{4m^2}{\pi}\int\,d^{2}\sigma\,\phi\,\pa_{+}x^{+}\pa_{-}x^{+}
\label{f2}
\eea
where the factor $2$  in the first line is obtained by replacing
$\psi^1\pa_{+}\eta^1$ with $\psi^2\pa_{+}\eta^2$. The interesting feature
of $F_{2}$ is that it can be described by the
one-loop triangle diagram shown in Fig.5.

%%%%%%% Fig. 5 %%%%%%%%%%%%%%%%
\begin{figure}[htbp]
\begin{center}
\includegraphics*[scale=.60]{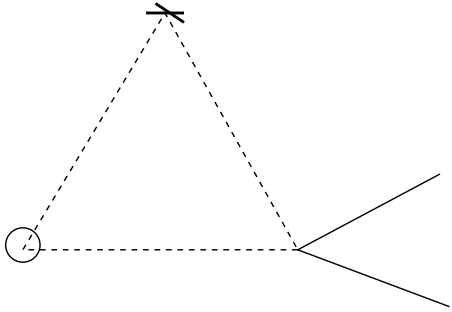}
\vspace{-0.7cm}
\end{center}
\caption{The triangle one-loop logarithmic divergence from
the contraction among $\psi^1\pa_{+}\eta^1$, 
$\psi^1\psi^2$, $\eta^1\eta^2$, where the
cross represents $\psi^1\psi^2$, the circle stands for $\psi^1\pa_{+}\eta^1
\epsilon\phi$ and the rest corresponds to $\eta^1\eta^2\pa_{+}x^+\pa_{-}x^+$.}
\end{figure}
%%%%%%%%%%%%%%%%%%%%%%%%%%%%%%%

The third one can be obtained from the contraction between 
the part of $F_{1}$ and $\pa_{+}x^{-}\pa_{-}x^{+}\epsilon\phi$
\bea
F_{3}&=&\frac{2}{(2\pi\a^{\prime})^3}\int\,d^{2+\epsilon}\sigma_{1}\,
d^{2+\epsilon}\sigma_{2}\,d^{2+\epsilon}\sigma_{3}\,
im\wick{12}{<1\psi^1(\sigma_{1})<2\psi^2(\sigma_{1})
im\,>1\eta^1\,(\sigma_{2})>2\eta^2(\sigma_{2})}\nn
&&\cdot
\pa_{+}x^+(\sigma_{2})\pa_{-}\,
\wick{1}{<1x^{+}(\sigma_{2})
\pa_{+}>1x^{-}}(\sigma_{3})\pa_{-}x^{+}(\sigma_{3})\epsilon\phi
\nn
&=&
\frac{-4m^2}{\pi}\int\,d^{2}\sigma\,\phi\,\pa_{+}x^{+}\pa_{-}x^{+}
\label{f3}
\eea
which is shown similarly as Fig.4.

The total fermionic $\phi$-dependent effective action
can be obtained the same way as in the bosonic
case with the similar structure of Fig.3, where Fig.4 and Fig.5
instead of Fig.1 and Fig.2 are used as the ``fundamental blocks''.
 
Eqs.(\ref{b1}), (\ref{b2}), (\ref{b3}), (\ref{f1}), (\ref{f2})
and (\ref{f3}) show that the $B_{1}+B_{2}+B_{3}+F_{1}+F_{2}+F_{2}=0$,
which indicates that the total $\phi$-dependent effective action vanishes.
We therefore conclude that the type IIB GS superstring in semi-light cone
gauge on the pp-wave RR background is described by a superconformal
field theory.

Here we emphasize that there are some higher-loop
diagrams, but they can be decomposed as the products of the 
one-loop diagrams. This is because the quartic interaction term
takes the special form ${x^{i}}^{2}\pa_{+}x^{+}\pa_{-}x^{+}$ and
$x^{+}$ can only contract with $x^{-}$, while there is no interacting
term quadratic in $x^{-}$.

Let us study the quantum counterterm in (\ref{fs}) instead of
the $\phi$-dependent effective action. The only logarithmic divergence
in bosonic part is 
\bea
b_{\epsilon}&=&\frac{1}{2\pi\a^{\prime}}\int\,d^{2+\epsilon}\sigma\,
\bigg(-\frac{1}{2}m^2\,\wick{1}{<1x^i\,>1x^i\,}
\pa_{+}x^{+}\pa_{-}x^{+}  
\bigg)\nn
&=&
-\frac{2m^2}{\pi\epsilon}\int\,d^{2}\sigma\,\pa_{+}x^{+}\pa_{-}x^{+}
\label{cb}
\eea
which can be illustrated by Fig.1.

The logarithmically divergent term in fermionic action is 
\bea
f_{\epsilon}&=&\frac{1}{(2\pi\a^{\prime})^2}\int\,d^{2+\epsilon}\sigma_{1}\,
d^{2+\epsilon}\sigma_{2}\,
im\wick{12}{<1\psi^1(\sigma_{1})<2\psi^2(\sigma_{1})
im\,>1\eta^1\,(\sigma_{2})>2\eta^2(\sigma_{2})}\nn
&&\cdot
\pa_{+}x^+(\sigma_{2})\pa_{-}\,
x^+(\sigma_{2})]\nn
&=&
\frac{2m^2}{\pi\epsilon}\int\,d^{2}\sigma\,\pa_{+}x^{+}\pa_{-}x^{+}
\label{cf}
\eea
which can be shown by Fig.4, but the cross represents the factor $im$.

Like the $\phi$-dependent effective action, the counterterms
in bosonic and fermionic parts are not zero respectively, but
the total counterterm for the action (\ref{fs}) vanishes:
$b_{\epsilon}+f_{\epsilon}=0$.

\section{Polygon divergent structure in group coordinates}
\label{s4}

The coordinates $(x^{\pm},x^i)$ are called harmonic coordinates
and are physically convenient in calculating the string spectrum. To display
the symmetries of the geometry and compute scattering 
amplitudes \cite{bit}, the so called ``group coordinates'' are more 
suitable \cite{gv, jn}. In group coodinates, the metric takes the 
form \cite{gv, jn}
\bea
ds^2\, = 2dy^+\,dy^-\, + e^{\pm\,2imy^+}d{y^i}^2
\label{me}
\eea
where the group coodinates are related to the harmonic coordinates by
\bea
x^+ &=& y^+\,,\nn
x^- &=& y^-\, \mp\frac{im}{2}e^{\pm\,2imy^+}y^i\,y^i\,,\nn
x^i &=& e^{\pm\,imy^+}\,y^i\,.
\eea
In \cite{gv, jn}, no criterion is given to determine the sign $\pm$.
Here we assign $\pm$ in the way that preserves conformal
invariance, and the metric is then
\bea
ds^2\, = 2dy^+\,dy^-\, + e^{2imy^+}\sum\limits^4_{i=1}\,d{y^i}^2
+ e^{-2imy^+}\sum\limits^8_{i'=5}\,d{y^{i'}}^2
\label{metric1}
\eea
with
\bea
x^+ &=& y^+\,,\nn
x^- &=& y^-\, -\frac{im}{2}e^{\,2imy^+}\sum\limits^4_{i=1}y^i\,y^i
+\frac{im}{2}e^{-2imy^+}\sum\limits^8_{i'=5}y^{i'}y^{i'},\nn
x^i &=& e^{\,imy^+}\,y^i\,\nn
x^{i'}&=&e^{-\,imy^+}y^{i'}
\label{trans}
\eea
where $i=1,\cdots,4$ and  $i'=5,\cdots,8$. 

Notice that our choice of signs has the property of preserving the volume
of the coordinate system.
The Jacobian $J(\frac{\pa\,x^{\mu}}{\pa\,y^{\nu}})$ gives
the measure factor $e^{i(4-4)m\,y^{+}}=1$. 
This assignment of $\pm$
also makes manifest the $SO(4)\times\,SO(4)$ symmetry.

Though the coordinates and metric are complex, 
the vertex operators in group coordinates
are quite simple \cite{gv, jn}. 
The massless scalar field
solution of the Laplacian \cite{lor} in harmonic coordinates is
\bea
V_{T}(x)=e^{ik_{+}x^{+}+ik_{-}x^{-}}\prod\limits_{j=1}^8\,e^{-\a_{j}
x_{j}^2\,/4}H_{n_j}(\sqrt{\frac{\a_{j}}{2}}x_{j})
\label{opeh}
\eea
with $k_{+}=\sum\limits_{j}\frac{\a_{j}}{k_{-}}(n_j\,+\frac{1}{2})$,
and $\a_{j}=\pm\,mk_{-}$. For $\a_{j}>0$, the $H_{n_j}$
are Hermite polynomials.
In group coordinates, the solution is simplified to \cite{gv, jn}
\bea
V_{T}(y)=e^{i\int\limits_{0}^{y^+}d{y^+}\,k_{+}({y^+})+ik_{-}y^{-}+
ik_{i}y^i\,}
\label{opeg}
\eea
where $k_{-}$, $k_{i}$ are constants and play the role of
components of the momentum.\footnote{After some
modification, the solution can be identified as tachyon vertex operator
in string theory \cite{jn}.}

In terms of group coordinates, the partition function (\ref{pf}) turns into
\beq
Z=\int\,Dy^{\mu}D\psi^{1}D\eta^{1}D\psi^{2}D\eta^{2}DbDc\,e^{-S}
\eeq
with
\bea
S&=&-\frac{1}{2\pi\a^{\prime}}\int\,d^{2}\sigma\,\bigg(
-y^{-}\pa_{+}\pa_{-}y^{+}
+ \frac{1}{2}\sum\limits^4_{i=1}\pa_{+}y^{i}\pa_{-}y^{i} 
+ \frac{1}{2}\sum\limits^8_{i'=5}\pa_{+}y^{i'}\pa_{-}y^{i'}
+ i\psi^1\,\pa_{+}\eta^1\,\nn
&&
+i\psi^2\,\pa_{-}\eta^2\, + {\cal\, L}_{bc}
+ \frac{1}{2}(e^{2imy^+}-1)\sum\limits^4_{i=1}\pa_{+}y^{i}\pa_{-}y^{i} 
+ \frac{1}{2}(e^{-2imy^+}-1)\sum\limits^8_{i'=5}\pa_{+}y^{i'}\pa_{-}y^{i'}\nn
&& 
+ i\,m\psi^1\,\psi^2\,
+ i\,m\eta^1\eta^2\,\pa_{+}y^{+}\pa_{-}y^{+}\bigg)
\label{ga}
\eea
Here the fermionic part is the same as that in harmonic coordinates,
and the fermionic quantum counterterm does not change in group coordinates,
so in group coordinates
we only need to consider the bosonic quantum counterterm.

The bosonic interaction is now changed into 
$\frac{1}{2}(e^{2imy^+}-1)\sum\limits^4_{i=1}\pa_{+}y^{i}\pa_{-}y^{i}
=-\frac{1}{2}(e^{2imy^+}-1)\sum\limits^4_{i=1}\pa_{a}y^{i}\pa_{a}y^{i}$, and
$-\frac{1}{2}(e^{-2imy^+}-1)\sum\limits^8_{i'=5}\pa_{a}y^{i'}\pa_{a}y^{i'}$. 
In group coordinates, the vertex operators and calculation
of scattering amplitudes are simple, but the action (\ref{ga})
and calculation of string spectrum are more complicated
than in harmonic coordinates. 

Though the $y^{i}$ interacts with $y^{i'}$ through $y^+$ and $y^-$,
when we calculate the bosonic quantum counterterm and $\phi$-dependent
effective action, $y^{i}$ still decouples effectively from $y^{i'}$.
Thus in the following calculation, we first consider $y^{i}$, then
add the contribution from $y^{i'}$ by replacing $m$ by $-m$.
Since quadratic divergences like the 
contact term $\delta\,(0)$ can be omitted in dimensional regularization,
the contraction between $y^{i}$ and $y^{i}$ does not create any counterterm.
To extract the logarithmic divergence, we expand
\bea
e^{2imy^+}(\sigma_{2})&=&e^{2imy^+}(\sigma_{1})+(\sigma_{2}-\sigma_{1})_a\cdot
\pa_{a}e^{2imy^+}(\sigma_{1})\nn
&&
+\frac{1}{2!}(\sigma_{2}-\sigma_{1})_a\,
(\sigma_{2}-\sigma_{1})_b\,\pa_{a}\pa_{b}e^{2imy^+}(\sigma_{1})+\cdots
\eea
where the higher-order terms only contribute to the finite term, so they
can be omitted.

The first logarithmic divergence comes from the contraction
\bea
{\cal A}_{2}&=&\frac{1}{2!}\cdot\,(-1)^2\,\cdot\,
(\frac{1}{2})^2\,\cdot\,2\cdot\frac{1}{(2\pi\a^{\prime})^2}
\int\,d^{2+\epsilon}\sigma_{1}\,
d^{2+\epsilon}\sigma_{2}\,(e^{2imy^+}-1)(\sigma_{1})
\cdot\,(e^{2imy^+}-1)(\sigma_{2})\nn
&&
\cdot\,i\wick{12}{\pa_{a}<1y^i(\sigma_{1})\pa_{a}<2y^i(\sigma_{1})
\pa_{b}>2y^j(\sigma_{2})\pa_{b}>1y^j(\sigma_{2})}\nn
&=&
-\frac{1}{4\pi\epsilon}\int\,d^{2}\sigma\,
(e^{2imy^+}-1)
\pa_{+}\pa_{-}e^{2imy^+}
\label{a2}
\eea
where the factor $2$ is from two sorts of contraction
between  $\pa_{a}y^i\pa_{a}y^i$ and  $\pa_{b}y^j\pa_{b}y^j$,
and the finite terms have been omitted. The Feynman diagram
for ${\cal A}_{2}$ is drawn in Fig.6.

%%%%%%% Fig. 6 %%%%%%%%%%%%%%%%
\begin{figure}[htbp]
\begin{center}
\includegraphics*[scale=.60]{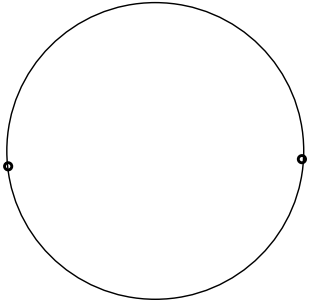}
\vspace{-0.7cm}
\end{center}
\caption{The points represent the factor $e^{2imy^+}-1$.}
\end{figure}
%%%%%%%%%%%%%%%%%%%%%%%%%%%%%%%

Besides ${\cal A}_{2}$, there are polygon contributions
to the logarithmic divergences. The $n$-gon logarithmic divergence
${\cal A}_{n}$ ($n=2,3,4,\cdots$) is
\bea
{\cal A}_{n}&=&\frac{1}{n!}\cdot\,(-1)^n\,\cdot\,
(\frac{1}{2})^n\,\cdot\,2^{n-1}\cdot\,(n-1)!\cdot\frac{1}{(2\pi\a^{\prime})^n}
\int\,d^{2+\epsilon}\sigma_{1}\,\cdots
d^{2+\epsilon}\sigma_{n}\,(e^{2imy^+}-1)(\sigma_{1})\nn
&&
\cdot\,(e^{2imy^+}-1)(\sigma_{2})\cdots\,(e^{2imy^+}-1)(\sigma_{n})\nn
&&
\cdot\,\wick{1234}{\pa_{a_1}<1y^{i_1}(\sigma_{1})
\pa_{a_1}<2y^{i_1}(\sigma_{1})
\pa_{a_2}>2y^{i_2}(\sigma_{2})
\pa_{a_2}<3y^{i_2}(\sigma_{2})>3\cdot\cdots
<4\cdot\,\pa_{a_n}>4y^{i_n}(\sigma_{n})
\pa_{a_n}>1y^{i_n}(\sigma_{n})}\nn
&=&
\frac{2(-1)^n}{(2\pi)^{2n}n}
\int\,d^{2+\epsilon}\sigma_1\cdots\,d^{2+\epsilon}\sigma_n
d^{2+\epsilon}k_1\cdots\,d^{2+\epsilon}k_n
\frac{(k_1)_{a_1}(k_1)_{a_2}}{k^2_1}
\frac{(k_2)_{a_2}(k_2)_{a_3}}{k^2_2}\nn
&&
\cdots\frac{(k_{n-1})_{a_{n-1}}(k_{n-1})_{a_n}}{k^2_{n-1}}
\frac{(k_{n})_{a_{n}}(k_{n})_{a_1}}{k^2_{n}}
\cdot\,e^{ik_1\cdot(\sigma_1\,-\sigma_2)}
\,e^{ik_2\cdot(\sigma_2\,-\sigma_3)}\cdots
\,e^{ik_{n-1}\cdot(\sigma_{n-1}\,-\sigma_n)}
\,e^{ik_{n}\cdot(\sigma_{n}\,-\sigma_1)}\nn
&&
\cdot
\bigg\{\frac{1}{2}(e^{2imy^+}-1)^{n-1}(\sigma_1)\bigg[\sum\limits^n_{i=2}
(\sigma_i\,-\sigma_1)^a\,(\sigma_i\,-\sigma_1)^b\,\bigg]\pa_a\pa_b
\,e^{2imy^+}(\sigma_1)\nn
&&
+(e^{2imy^+}-1)^{n-2}(\sigma_1)\bigg[\sum\limits^{n-1}_{i=2}
\sum\limits^{n}_{j=i+1}(\sigma_i\,-\sigma_1)^a\,(\sigma_j\,-\sigma_1)^b\,
\bigg]\pa_a\,e^{2imy^+}\pa_b\,e^{2imy^+}(\sigma_1)\bigg\}
\label{an}
\eea
where the quadratic divergence $\sim\delta(0)$ and finite terms have
been omitted. The Feynman diagram for this n-gon contribution
is shown in Fig.7.

%%%%%%% Fig. 7 %%%%%%%%%%%%%%%%
\begin{figure}[htbp]
\begin{center}
\includegraphics*[scale=.60]{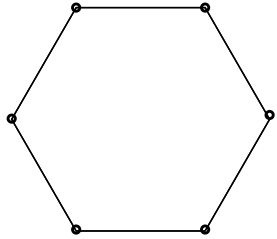}
\vspace{-0.7cm}
\end{center}
\caption{The n(=6)-gon logarithmic divergence where the circles stand
for the factor $e^{2imy^+}-1$.}
\end{figure}
%%%%%%%%%%%%%%%%%%%%%%%%%%%%%%%

The terms like $(\sigma_i\,-\sigma_1)^a\,(\sigma_j\,-\sigma_k)^b$
with $j>k\geq\,i+1$ vanish after integration, so that (\ref{an}) is reduced to
\bea
{\cal A}_{n}&=&\frac{2(-1)^n}{(2\pi)^{2n}n}
\int\,d^{2+\epsilon}\sigma_1\cdots\,d^{2+\epsilon}\sigma_n
d^{2+\epsilon}k_1\cdots\,d^{2+\epsilon}k_n
\frac{(k_1)_{a_1}(k_1)_{a_2}}{k^2_1}
\frac{(k_2)_{a_2}(k_2)_{a_3}}{k^2_2}\nn
&&
\cdots\frac{(k_{n-1})_{a_{n-1}}(k_{n-1})_{a_n}}{k^2_{n-1}}
\frac{(k_{n})_{a_{n}}(k_{n})_{a_1}}{k^2_{n}}
\cdot\,e^{ik_1\cdot(\sigma_1\,-\sigma_2)}
\,e^{ik_2\cdot(\sigma_2\,-\sigma_3)}\cdots
\,e^{ik_{n-1}\cdot(\sigma_{n-1}\,-\sigma_n)}
\,e^{ik_{n}\cdot(\sigma_{n}\,-\sigma_1)}\nn
&&
\cdot
\bigg\{\frac{1}{2}(e^{2imy^+}-1)^{n-1}(\sigma_1)\bigg[
(\sigma_n\,-\sigma_1)^a(\sigma_n\,-\sigma_1)^b+
\sum\limits^{n-1}_{i=2}
(\sigma_i\,-\sigma_1)^a\,(\sigma_i\,-\sigma_1)^b\,\bigg]\nn
&&
\cdot\pa_a\pa_b
\,e^{2imy^+}(\sigma_1)
+(e^{2imy^+}-1)^{n-2}(\sigma_1)\bigg[
-(n-2)(\sigma_2\,-\sigma_1)^a\,(\sigma_1\,-\sigma_n)^b +\nn
&&
\sum\limits^{n-2}_{i=2}
(n-1-i)(\sigma_i\,-\sigma_1)^a\,(\sigma_{i+1}\,
-\sigma_1)^b\,
\bigg]\pa_a\,e^{2imy^+}\pa_b\,e^{2imy^+}(\sigma_1)\bigg\}.
\label{an1}
\eea
The terms $(\sigma_i\,-\sigma_1)^a\,(\sigma_{i+1}\,-\sigma_1)^b$
and $(\sigma_i\,-\sigma_1)^a\,(\sigma_{i}\,-\sigma_1)^b$ can
be rewritten as 
\bea
(\sigma_i\,-\sigma_1)^a\,(\sigma_{i+1}\,-\sigma_1)^b&=&
\sum\limits^{i}_{l=2}(\sigma_l\,-\sigma_{l-1})^a\,(\sigma_{l}\,
-\sigma_{l-1})^b + \sum\limits^{i}_{l=2}(\sigma_l\,-\sigma_{l-1})^a\,
(\sigma_{l+1}\,-\sigma_{l})^b\nn
&& 
+ \sum\limits^{i}_{l=3}
(\sigma_l\,-\sigma_{l-1})^a\,(\sigma_{l-1}\, -\sigma_{l-2})^b + 
\sum\limits_{j\neq\,l-1,l,l+1}
(\sigma_l\,-\sigma_{l-1})^a\,(\sigma_{j}\, -\sigma_{j-1})^b\nn
(\sigma_i\,-\sigma_1)^a\,(\sigma_{i}\,-\sigma_1)^b&=&
\sum\limits^{i}_{l=2}(\sigma_l\,-\sigma_{l-1})^a\,(\sigma_{l}\,
-\sigma_{l-1})^b + \sum\limits^{i-1}_{l=2}(\sigma_l\,-\sigma_{l-1})^a\,
(\sigma_{l+1}\,-\sigma_{l})^b\nn
&& 
+ \sum\limits^{i}_{l=3}
(\sigma_l\,-\sigma_{l-1})^a\,(\sigma_{l-1}\, -\sigma_{l-2})^b\nn
&&  
+ \sum\limits_{j\neq\,l-1,l,l+1}
(\sigma_l\,-\sigma_{l-1})^a\,(\sigma_{j}\, -\sigma_{j-1})^b
\eea
After integration, only the terms like 
$(\sigma_l\,-\sigma_{l-1})^a\,(\sigma_{j}\, -\sigma_{j-1})^b$
with $j=l-1, l, l+1$ are not zero, so we have 
\bea
{\cal A}_{n}&=&\frac{2(-1)^n}{(2\pi)^{2n}n}
\int\,d^{2+\epsilon}\sigma_1\cdots\,d^{2+\epsilon}\sigma_n
d^{2+\epsilon}k_1\cdots\,d^{2+\epsilon}k_n
\frac{(k_1)_{a_1}(k_1)_{a_2}}{k^2_1}
\frac{(k_2)_{a_2}(k_2)_{a_3}}{k^2_2}\nn
&&
\cdots\frac{(k_{n-1})_{a_{n-1}}(k_{n-1})_{a_n}}{k^2_{n-1}}
\frac{(k_{n})_{a_{n}}(k_{n})_{a_1}}{k^2_{n}}
\cdot\,e^{ik_1\cdot(\sigma_1\,-\sigma_2)}
\,e^{ik_2\cdot(\sigma_2\,-\sigma_3)}\cdots
\,e^{ik_{n-1}\cdot(\sigma_{n-1}\,-\sigma_n)}
\,e^{ik_{n}\cdot(\sigma_{n}\,-\sigma_1)}\nn
&&
\cdot
\bigg\{\frac{1}{2}(e^{2imy^+}-1)^{n-1}(\sigma_1)\bigg[
(\sigma_n\,-\sigma_1)^a(\sigma_n\,-\sigma_1)^b+
\sum\limits^{n-1}_{i=2}\bigg(\sum\limits^{i}_{l=2}
(\sigma_l\,-\sigma_{l-1})^a\,(\sigma_l\,-\sigma_{l-1})^b+\nn
&&
\sum\limits^{i-1}_{l=2}
(\sigma_l\,-\sigma_{l-1})^a\,(\sigma_{l+1}\,-\sigma_{l})^b+
\sum\limits^{i}_{l=3}
(\sigma_l\,-\sigma_{l-1})^a\,(\sigma_{l-1}\,-\sigma_{l-2})^b
\,\bigg)\bigg]
\cdot\pa_a\pa_b
\,e^{2imy^+}(\sigma_1)\nn
&&
+(e^{2imy^+}-1)^{n-2}(\sigma_1)\bigg[
-(n-2)(\sigma_2\,-\sigma_1)^a\,(\sigma_1\,-\sigma_n)^b +
\sum\limits^{n-2}_{i=2}(n-1-i)\nn
&&
\cdot\bigg(\sum\limits^{i}_{l=2}
(\sigma_l\,-\sigma_{l-1})^a\,(\sigma_l\,-\sigma_{l-1})^b+
\sum\limits^{i}_{l=2}
(\sigma_l\,-\sigma_{l-1})^a\,(\sigma_{l+1}\,-\sigma_{l})^b+\nn
&&
\sum\limits^{i}_{l=3}
(\sigma_l\,-\sigma_{l-1})^a\,(\sigma_{l-1}\,-\sigma_{l-2})^b
\,\bigg)\bigg]
\cdot\pa_a\,e^{2imy^+}\pa_b\,e^{2imy^+}(\sigma_1).
\label{an2}
\eea

To further calculate (\ref{an2}), we consider the typical term $P_l$
\bea
P_l&=&\frac{2(-1)^n}{(2\pi)^{2n}n}
\int\,d^{2+\epsilon}\sigma_1\cdots\,d^{2+\epsilon}\sigma_n
d^{2+\epsilon}k_1\cdots\,d^{2+\epsilon}k_n
\frac{(k_1)_{a_1}(k_1)_{a_2}}{k^2_1}
\frac{(k_2)_{a_2}(k_2)_{a_3}}{k^2_2}\nn
&&
\cdots\frac{(k_{n-1})_{a_{n-1}}(k_{n-1})_{a_n}}{k^2_{n-1}}
\frac{(k_{n})_{a_{n}}(k_{n})_{a_1}}{k^2_{n}}
\cdot\,e^{ik_1\cdot(\sigma_1\,-\sigma_2)}
\,e^{ik_2\cdot(\sigma_2\,-\sigma_3)}\cdots
\,e^{ik_{n-1}\cdot(\sigma_{n-1}\,-\sigma_n)}
\,e^{ik_{n}\cdot(\sigma_{n}\,-\sigma_1)}\nn
&&
(\sigma_l\,-\sigma_{l-1})^a\,(\sigma_{l+1}\,-\sigma_{l})^b\cdot\,
F_{ab}(\sigma_1)\nn
&=&
\frac{2(-1)^n}{(2\pi)^{2n}n}
\int\,d^{2+\epsilon}\sigma_1\cdots\,d^{2+\epsilon}\sigma_n
d^{2+\epsilon}k_1\cdots\,d^{2+\epsilon}k_n\,F_{ab}(\sigma_1)
\frac{(k_1)_{a_1}(k_1)_{a_2}}{k^2_1}
\frac{(k_2)_{a_2}(k_2)_{a_3}}{k^2_2}\nn
&&
\cdots\frac{(k_{n-1})_{a_{n-1}}(k_{n-1})_{a_n}}{k^2_{n-1}}
\frac{(k_{n})_{a_{n}}(k_{n})_{a_1}}{k^2_{n}}
\cdot\,e^{ik_1\cdot(\sigma_1\,-\sigma_2)}
\,e^{ik_2\cdot(\sigma_2\,-\sigma_3)}\cdots\nn
&&
e^{ik_{l-2}\cdot(\sigma_{l-2}\,-\sigma_{l-1})}
\bigg[\frac{\pa}{i\pa(k_{l-1})_a}e^{ik_{l-l}\cdot(\sigma_{l-1}\,-\sigma_{l})} 
\bigg]\cdot
\bigg[\frac{\pa}{i\pa(k_{l})_b}
e^{ik_{l}\cdot(\sigma_{l}\,-\sigma_{l+1})} 
\bigg]
e^{ik_{l+1}\cdot(\sigma_{l+1}\,-\sigma_{l+2})}\nn
&& 
\cdots
\,e^{ik_{n-1}\cdot(\sigma_{n-1}\,-\sigma_n)}
\,e^{ik_{n}\cdot(\sigma_{n}\,-\sigma_1)}.
\label{type}
\eea
First integrating $k_{l-1}$ and $k_{l}$ by parts, then
integrating out $\sigma_2,\cdots,\sigma_n$,
$k_1, k_{l-1}, k_{l+1},\cdots, k_n$, and renaming
$\sigma_1\rightarrow\sigma$, $ k_{l}\rightarrow\,k$, 
we have
\bea
P_l&=&-\frac{2(-1)^n}{(2\pi)^{2n}n}
\int\,d^{2+\epsilon}\sigma\,d^{2+\epsilon}k
\frac{\pa}{\pa\,k_a}(\frac{k_{a_1}\,k_{a_2}}{k^2})\cdot
\frac{\pa}{\pa\,k_b}(\frac{k_{a_2}\,k_{a_3}}{k^2})
\frac{k_{a_3}\,k_{a_1}}{k^2}F_{ab}(\sigma)\nn
&=& 
-\frac{(-1)^n}{2n\pi\,\epsilon}\int\,d^{2}\sigma\,F_{ab}(\sigma)\delta^{ab}
\label{pl}
\eea
where $P_l$ is independent of the index $l$. 

Similarly, the other typical term $Q_l$ is
\bea
Q_l&=&\frac{2(-1)^n}{(2\pi)^{2n}n}
\int\,d^{2+\epsilon}\sigma_1\cdots\,d^{2+\epsilon}\sigma_n
d^{2+\epsilon}k_1\cdots\,d^{2+\epsilon}k_n
\frac{(k_1)_{a_1}(k_1)_{a_2}}{k^2_1}
\frac{(k_2)_{a_2}(k_2)_{a_3}}{k^2_2}\nn
&&
\cdots\frac{(k_{n-1})_{a_{n-1}}(k_{n-1})_{a_n}}{k^2_{n-1}}
\frac{(k_{n})_{a_{n}}(k_{n})_{a_1}}{k^2_{n}}
\cdot\,e^{ik_1\cdot(\sigma_1\,-\sigma_2)}
\,e^{ik_2\cdot(\sigma_2\,-\sigma_3)}\cdots
\,e^{ik_{n-1}\cdot(\sigma_{n-1}\,-\sigma_n)}
\,e^{ik_{n}\cdot(\sigma_{n}\,-\sigma_1)}\nn
&&
(\sigma_l\,-\sigma_{l-1})^a\,(\sigma_{l}\,-\sigma_{l-1})^b\cdot\,
F_{ab}(\sigma_1)\nn
&=&
\frac{(-1)^n}{n\pi\,\epsilon}\int\,d^{2}\sigma\,F_{ab}(\sigma)\delta^{ab}.
\label{ql}
\eea

Inserting (\ref{pl}) and (\ref{ql}) into (\ref{an2}), we obtain
\bea
{\cal A}_n\,&=&\frac{(-1)^n}{2n\pi\epsilon}\int\,d^2\sigma
\bigg\{\frac{1}{2}\bigg[2+
\sum\limits^{n-1}_{i=2}\bigg(2(i-1)-(i-2)-(i-2)\bigg)\bigg](e^{2imy^+}-1)^{n-1}
\pa_a\pa_a\,e^{2imy^+}\nn
&&
+\bigg[(n-2)+\sum\limits^{n-2}_{i=2}
(n-1-i)\bigg(2(i-1)-(i-1)-(i-2)\bigg)\bigg](e^{2imy^+}-1)^{n-2}
\pa_a\,e^{2imy^+}\pa_a\,e^{2imy^+}\bigg\}\nn
&=&
\frac{(-1)^{n-2}}{2\pi\epsilon}\frac{1}{n}\int\,d^2\sigma
\bigg\{\frac{(n-1)(n-2)}{2}(e^{2imy^+}-1)^{n-2}
\pa_a\,e^{2imy^+}\pa_a\,e^{2imy^+}\nn
&&
+(n-1)(e^{2imy^+}-1)^{n-1}
\pa_a\pa_a\,e^{2imy^+}\bigg\}.
\label{coun}
\eea
After integration by parts, ${\cal A}_n$ is reduced to
\bea
{\cal A}_n\,=
\frac{(-1)^{n-2}}{4\pi\epsilon}\int\,d^2\sigma
(e^{2imy^+}-1)^{n-1}
\pa_a\pa_a\,e^{2imy^+}
\eea
which is the logarithmically divergent term from the n-gon.

The total logarithmically divergent part ${\cal A}$ from $y^i$ is
\bea
{\cal A}=\sum\limits^\infty_{n=2}{\cal A}_n
=-\frac{m^2}{\pi\epsilon}\int\,d^2\sigma
\pa_+\,{y^+}\pa_-\,{y^+}.
\eea
Thus the total logarithmically divergent term from $y^i$ and
$y^{i'}$ is obtained by adding the same term, replacing $m$ by $-m$
\bea
{\cal A}_T=-\frac{2m^2}{\pi\epsilon}\int\,d^2\sigma
\pa_+\,{y^+}\pa_-\,{y^+}
\eea
This is exactly the same as in harmonic coordinates (\ref{cb}), 
and can be illustrated by Fig.8.

%%%%%%% Fig. 8 %%%%%%%%%%%%%%%%
\begin{figure}[htbp]
\begin{center}
\includegraphics*[scale=.60]{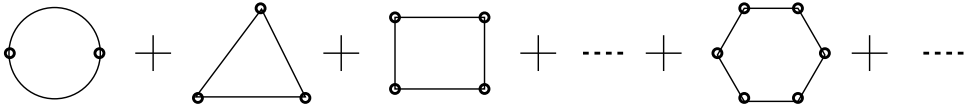}
\vspace{-0.7cm}
\end{center}
\caption{The total logarithmic divergence from
the sum of the n-gons.}
\end{figure}
%%%%%%%%%%%%%%%%%%%%%%%%%%%%%%%

Now let us consider the $\phi$-dependent effective action
in group coordinates. When $\epsilon$ is small, the dimensional
continuation of the action (\ref{ga}) is
\bea
S&=&-\frac{1}{2\pi\a^{\prime}}\int\,d^{2+\epsilon}\sigma\,\bigg(
-y^{-}\pa_{+}\pa_{-}y^{+}
+ \frac{1}{2}\sum\limits^4_{i=1}\pa_{+}y^{i}\pa_{-}y^{i} 
+ \frac{1}{2}\sum\limits^8_{i'=5}\pa_{+}y^{i'}\pa_{-}y^{i'}\nn
&&
+ i\psi^1\,\pa_{+}\eta^1\,+i\psi^2\,\pa_{-}\eta^2\, + {\cal\, L}_{bc}
-y^{-}\pa_{+}\pa_{-}y^{+}\epsilon\phi +  
 \frac{1}{2}\sum\limits^4_{i=1}\pa_{+}y^{i}\pa_{-}y^{i}\epsilon\phi \nn
&&
+ \frac{1}{2}\sum\limits^8_{i'=5}\pa_{+}y^{i'}\pa_{-}y^{i'}\epsilon\phi
+i\psi^1\,\pa_{+}\eta^1\,\epsilon\phi
+i\psi^2\,\pa_{-}\eta^2\,\epsilon\phi
+ {\cal\, L}_{bc}\,\epsilon\phi\nn
&&
+\frac{1}{2}(e^{2imy^+}-1)\sum\limits^4_{i=1}\pa_{+}{y^{i}}\pa_{+}y^{i}
+\frac{1}{2}(e^{2imy^+}-1)\sum\limits^4_{i=1}
\pa_{+}{y^{i}}\pa_{+}y^{i}\epsilon\phi\nn
&&
+\frac{1}{2}(e^{-2imy^+}-1)\sum\limits^8_{i'=5}\pa_{+}{y^{i'}}\pa_{+}y^{i'}
+\frac{1}{2}(e^{-2imy^+}-1)\sum\limits^8_{i'=5}
\pa_{+}{y^{i'}}\pa_{+}y^{i'}\epsilon\phi
+ i\,m\psi^1\,\psi^2\,\nn
&&
+ i\,m\psi^1\,\psi^2\,\epsilon\phi
+ i\,m\eta^1\eta^2\,\pa_{+}y^{+}\pa_{-}y^{+}
+ i\,m\eta^1\eta^2\,\pa_{+}y^{+}\pa_{-}y^{+}\epsilon\phi
\bigg)
\label{ga1}
\eea
where the relevant bosonic interactive Lagrangian is
$\frac{1}{2}\pa_{+}y^{i}\pa_{-}y^{i}\epsilon\phi
+\frac{1}{2}(e^{2imy^+}-1)\pa_{+}{y^{i}}\pa_{+}y^{i}
+\frac{1}{2}(e^{2imy^+}-1)\pa_{+}{y^{i}}\pa_{+}y^{i}\epsilon\phi$, similar
for $y^{i'}$.

Following the procedure of the calculation of the logarithmic
divergence and exploiting (\ref{coun}), the first part of the
$\phi$-dependent effective action from n-gon is 
\bea
{\cal L}_{(n)}^{\phi}\,&=&
\frac{(-1)^{n-2}}{2\pi}\int\,d^2\sigma
\bigg\{\frac{(n-1)(n-2)}{2}\phi\,(e^{2imy^+}-1)^{n-3}
\pa_a\,e^{2imy^+}\pa_a\,e^{2imy^+}\nn
&&
+(n-1)\phi\,(e^{2imy^+}-1)^{n-2}
\pa_a\pa_a\,e^{2imy^+}\nn
&&
+\frac{(n-1)(n-2)}{2}\phi\,(e^{2imy^+}-1)^{n-2}
\pa_a\,e^{2imy^+}\pa_a\,e^{2imy^+}\nn
&&
+(n-1)\phi\,(e^{2imy^+}-1)^{n-1}
\pa_a\pa_a\,e^{2imy^+}\bigg\}
\eea
where the first two terms are obtained by replacing one point of
Fig.6 $(e^{2imy^+}-1)$ by $\epsilon\phi$, and the last two
terms by replacing one point of
Fig.6 $(e^{2imy^+}-1)$ by $(e^{2imy^+}-1)\epsilon\phi$.
In deriving the quantum counterterm (\ref{coun}),
the only interacting term is 
$-\frac{1}{2}(e^{2imy^+}-1)\pa_{a}{y^{i}}\pa_{a}y^{i}$. However,
when calculating the $\phi$-dependent effective Lagrangian from n-gon,
there are three interacting terms 
$-\frac{1}{2}(e^{2imy^+}-1)\pa_{a}{y^{i}}\pa_{a}y^{i}$,
$-\frac{1}{2}(e^{2imy^+}-1)\pa_{a}{y^{i}}\pa_{a}y^{i}\epsilon\phi$
and $-\frac{1}{2}\pa_{a}{y^{i}}\pa_{a}y^{i}\epsilon\phi$,
and the interacting terms 
$-\frac{1}{2}\pa_{a}{y^{i}}\pa_{a}y^{i}\epsilon\phi$ and 
$-\frac{1}{2}(e^{2imy^+}-1)\pa_{a}{y^{i}}\pa_{a}y^{i}\epsilon\phi$
only appear once respectively because of the logarithmic divergence.

The first part of the $\phi$-dependent effective action from $y^i$ is
\bea
{\cal L}^{\phi}\,=\sum\limits^\infty_{n=2}{\cal L}^{\phi}_{(n)}\,
=-\frac{mi}{\pi}\int\,d^2\sigma\phi
\pa_+\,\pa_{-}\,{y^+}
\eea
which is proportional to the equation of motion $\pa_+\,\pa_{-}\,{y^+}=0$,
and can be ignored in the process of renormalization.

Similar to $B_3$ and $F_3$ the second part of the 
$\phi$-dependent effective action which can be obtained by the
contraction between ${\cal A}$ and $\pa_{+}y^{-}\pa_{-}y^{+}\epsilon\phi$
is
\bea
\Delta{\cal L}^{\phi}\,&=&-\frac{2m^2}{2\pi\a^{\prime}}\int\,
d^{2+\epsilon}\sigma_{1}\,
d^{2+\epsilon}\sigma_{2}\,
\pa_{+}y^{+}(\sigma_{1})\pa_{-}\wick{1}{<1y^{+}(\sigma_{1}) 
)\pa_{+}>1y^{-}}(\sigma_{2})\pa_{-}y^{+}(\sigma_{2})\epsilon\phi\nn
&=&
\frac{2m^2}{\pi}\int\,d^{2}\sigma\,\phi\,\pa_{+}y^{+}\pa_{-}y^{+}.
\label{del}
\eea
The total logarithmically divergent term from $y^i$ and
$y^{i'}$ is obtained by adding the same term with $m$ replaced by $-m$
\bea
{\cal L}_T^{\phi}\,&=&-\frac{mi}{\pi}\int\,d^2\sigma\phi
\pa_+\,\pa_{-}\,{y^+}
+\frac{2m^2}{\pi}\int\,d^{2}\sigma\,\phi\,\pa_{+}y^{+}\pa_{-}y^{+}\nn
&&
-\frac{-mi}{\pi}\int\,d^2\sigma\phi
\pa_+\,\pa_{-}\,{y^+}
+\frac{2m^2}{\pi}\int\,d^{2}\sigma\,\phi\,\pa_{+}y^{+}\pa_{-}y^{+}\nn
&=&
\frac{4m^2}{\pi}\int\,d^{2}\sigma\,\phi\,\pa_{+}y^{+}\pa_{-}y^{+}
\label{el}
\eea
This cancels with the fermionic $\phi$-dependent effective action,
i.e., ${\cal L}_T^{\phi}+F_1\,+F_2\,+F_3\,=0$.
Thus the pp-wave GS superstring in group coordinates
is a conformal field theory.

\section{Summary and conclusion}
\label{s5}

We have studied the pp-wave
GS superstring in the semi-light cone gauge $g_{ab}=e^{2\phi}\delta_{ab}$,
$\bar{\gamma}^+\theta=0$. The original GS superstring
action with $SO(8)$ spinors has been recast into a simple form with
two $SU(4)$ spinors. For the $m=0$ case,
the conformal anomaly from $SU(4)$ spinors
has a coefficient $+8$ while the $x^{\mu}$
contribute $\frac{10}{2}$ and reparametrization ghost contribute
$-\frac{26}{2}$, thus the total conformal anomaly in (\ref{fs}) vanishes.

For the $m\neq 0$ case,
we have calculated the $\phi$-dependent bosonic
effective action in harmonic coordinates. 
When we compute the fermionic $\phi$-dependent 
effective action, we have found a new triangular one-loop Feynman diagram.
We have shown that the bosonic
$\phi$-dependent effective action cancels with the fermionic one, 
which indicates that the pp-wave GS superstring is a
exact conformal field theory. 
The quartic interacting term in pp-wave background
is ${x^{i}}^{2}\pa_{+}x^{+}\pa_{-}x^{+}$ and
$x^{+}$ can only contract with $x^{-}$, and there is no interacting
term quadratic in $x^{-}$, thus the higher-loop diagrams can
be decomposed as the products of the one-loop diagrams,
which is a crucial feature when we discuss the mixing
of the vertex operators.

We have introduced the group coordinates preserving $SO(4)\times\,SO(4)$
and conformal symmetry. And we found that in group coordinates  there are
logarithmic divergences from n-gons whose divergent structure
is more complicated than that in harmonic coordinates.
After summing over all contributions from n-gons, we have shown that
in group coordinates, the GS superstring on pp-wave RR 
background is also a conformal field theory.

In the above, to prove the conformal invariance
of the pp-wave GS superstring action, we have shown how the 
$\phi$-dependent effective action vanishes. The authors of
 \cite{berk,berkm,tse},
only commented on the UV finiteness of the string action.
When we construct the vertex operators on pp-wave, 
we will calculate one-particle irreducible diagrams of 
an insertion of a vertex operator. In this case, 
the method for calculating $\phi$-dependent effective action is
more useful than that for calculating UV finiteness of the string action.
Actually, our work has paved the way to construct 
the vertex operators and discuss the mixing 
of the  vertex operators on the pp-wave.

\begin{acknowledgments}
We thank N. Berkovits and P. Mathieu for their
helpful correspondence. This work was supported in part by NSERC.
\end{acknowledgments}

\newcommand{\bib}{\bibitem}
\newcommand{\NP}[1]{Nucl.\ Phys.\ {\bf #1}}
\newcommand{\AP}[1]{Ann.\ Phys.\ {\bf #1}}
\newcommand{\PL}[1]{Phys.\ Lett.\ {\bf #1}}
\newcommand{\CQG}[1]{Class. Quant. Gravity {\bf #1}}
\newcommand{\CMP}[1]{Comm.\ Math.\ Phys.\ {\bf #1}}
\newcommand{\PR}[1]{Phys.\ Rev.\ {\bf #1}}
\newcommand{\PRL}[1]{Phys.\ Rev.\ Lett.\ {\bf #1}}
\newcommand{\PRE}[1]{Phys.\ Rep.\ {\bf #1}}
\newcommand{\PTP}[1]{Prog.\ Theor.\ Phys.\ {\bf #1}}
\newcommand{\PTPS}[1]{Prog.\ Theor.\ Phys.\ Suppl.\ {\bf #1}}
\newcommand{\MPL}[1]{Mod.\ Phys.\ Lett.\ {\bf #1}}
\newcommand{\IJMP}[1]{Int.\ Jour.\ Mod.\ Phys.\ {\bf #1}}
\newcommand{\JHEP}[1]{J.\ High\ Energy\ Phys.\ {\bf #1}}

\end{document}